\documentclass{PoS}

\title{Non-Abelian magnetic monopoles responsible for quark confinement 
}

\ShortTitle{Non-Abelian magnetic monopoles}

\author{\speaker{Kei-Ichi Kondo}
\\
Department of Physics,  
Graduate School of Science, 
Chiba University, Chiba 263-8522, Japan
\\
        E-mail: \email{kondok@faculty.chiba-u.jp}}

\author{Akihiro Shibata
\\
Computing Research Center, High Energy Accelerator Research Organization,  Tsukuba  305-0801, Japan
\\
        E-mail: \email{akihiro.shibata@kek.jp}}

\author{Toru Shinohara
\\Department of Physics,  
Graduate School of Science, 
Chiba University, Chiba 263-8522, Japan
\\
        E-mail: \email{sinohara@graduate.chiba-u.jp}}

\author{Seikou Kato
\\
Fukui National College of Technology, Sabae 916-8507, Japan
\\
        E-mail: \email{skato@fukui-nct.ac.jp}}

\abstract{We show that the non-Abelian magnetic monopole defined in a gauge-invariant way in SU(3) Yang-Mills theory gives a dominant contribution to confinement of the fundamental quark, in sharp contrast to the SU(2) case. 
}

\FullConference{The many faces of QCD\\
		November 2-5, 2010\\
		Gent Belgium}

\usepackage{amsmath,amssymb}
\usepackage{mathrsfs}
\usepackage{bm}

\begin{document}

\section{Introduction}

The dual superconductor picture proposed long ago  \cite{dualsuper}  is believed to be a promising mechanics for quark confinement. 
For this mechanism to work, however, magnetic monopoles and their condensation are indispensable to cause the dual Meissner effect leading to the linear potential between quark and antiquark, namely,  
  area law of the Wilson loop average.  
The Abelian projection method proposed by 't Hooft \cite{tHooft81} can be used to introduce such magnetic monopoles into the pure Yang-Mills theory even without matter fields. 
Indeed, numerical evidences supporting the dual superconductor picture resulting from such magnetic monopoles have been accumulated since 1990  in pure SU(2) Yang-Mills theory \cite{SY90,SNW94,AS99}.
However, {\it the Abelian projection method explicitly breaks both the local gauge symmetry and the global color symmetry} by partial gauge  fixing from an original non-Abelian gauge group $G=SU(N)$ to the maximal torus subgroup, $H=U(1)^{N-1}$. 
Moreover, the Abelian dominance \cite{SY90} and  magnetic monopole dominance \cite{SNW94} were observed only in a special class of gauges, e.g., the maximally Abelian (MA) gauge and Laplacian Abelian (LA) gauge, realizing the idea of  Abelian projection.

For $G=SU(2)$, we have already succeeded to settle the issue of  gauge (in)dependence by {\it introducing  a gauge-invariant magnetic monopole in a gauge independent way}, based on another method: a non-Abelian Stokes theorem for the Wilson loop operator \cite{DP89,Kondo98b} and a new reformulation of Yang-Mills theory rewritten in terms of new field variables \cite{KMS06,KMS05,Kondo06} and \cite{KKMSSI05,IKKMSS06,SKKMSI07}, elaborating the technique proposed by Cho \cite{Cho80} and Duan and Ge \cite{DG79} independently, and later readdressed by Faddeev and Niemi \cite{FN99}.

For $G=SU(N)$, $N \ge 3$, there are no inevitable reasons why degrees of freedom associated with the maximal torus subgroup should be most dominant for quark confinement.
In this case, the problem is not settled yet. 
In this talk, 
we give a theoretical framework for describing {\it non-Abelian  dual superconductivity} in $D$-dimensional $SU(N)$ Yang-Mills theory, which should be compared with the conventional Abelian $U(1)^{N-1}$ dual superconductivity in $SU(N)$ Yang-Mills theory, hypothesized by  Abelian projection. 
We demonstrate that {\it an effective low-energy description for quarks in the fundamental representation} (abbreviated to rep. hereafter) {\it can be given by a set of non-Abelian restricted field variables} and that {\it non-Abelian $U(N-1)$ magnetic monopoles} in the sense of Goddard--Nuyts--Olive--Weinberg \cite{nAmm} {\it are the most dominant topological configurations for quark confinement} as conjectured in \cite{KT99,Kondo99Lattice99}.


\section{Wilson loop and gauge-inv. magnetic monopole}

A version of a non-Abelian Stokes theorem (NAST) for the Wilson loop operator originally invented by Diakonov and Petrov \cite{DP89} for $G=SU(2)$ was proved to hold \cite{Kondo98b} and was extended to $G=SU(N)$ \cite{KT99,Kondo08} in a unified way \cite{Kondo08} as a path-integral rep. by making use of a coherent state for the Lie group.  
For the Lie algebra $su(N)$-valued Yang-Mills field $\mathscr{A}_\mu(x)=\mathscr{A}_\mu^A(x) T_A$ with $su(N)$ generators $T_A$ ($A=1, \cdots, N^2-1$), 
the NAST enables one to rewrite a non-Abelian Wilson loop operator \begin{align}
  W_C[\mathscr{A}]  
:=& {\rm tr} \left[ \mathscr{P} \exp \left\{ ig_{\rm YM} \oint_{C} dx^\mu \mathscr{A}_\mu(x) \right\} \right]/{\rm tr}({\bf 1})  
 ,
\end{align}
 in terms of an Abelian-like potential $A_\mu$ as
\begin{equation}
 W_C[\mathscr{A}] = \int d\mu_{C}(g) \exp \left[  ig_{\rm YM} \oint_{C} A \right] ,
 \label{pre-NAST}
\end{equation}
where $g_{\rm YM}$ is the Yang-Mills coupling constant,  
  $d\mu_{C}(g):=\prod_{x \in C} d\mu(g_{x})$ with an invariant measure $d\mu$ on $G$  normalized as $\int d\mu(g_{x})=1$, $g_{x}$ is an element of a gauge group $G$ (more precisely, rep. $D_R(g_{x})$ of  $G$), and the one-form $A := A_\mu(x) dx^\mu$ is defined by
\begin{equation}
A_\mu(x) = {\rm tr}\{ \rho[ g_{x}^\dagger \mathscr{A}_\mu(x) g_{x} + ig_{\rm YM}^{-1} g_{x}^\dagger \partial_\mu g_{x} ] \} , \ g_{x} \in G .
\end{equation}
Here $\rho$ is defined as $\rho :=  | \Lambda \rangle \langle \Lambda |$ using a reference state (highest or lowest  weight state of the rep.) $| \Lambda  \rangle$ making a rep. of the Wilson loop we consider. 
Note that ${\rm tr}(\rho) = \langle \Lambda | \Lambda \rangle = 1$ follows from the normalization of $| \Lambda \rangle$. 
Then it is rewritten into the surface-integral form using a usual Stokes theorem:
\begin{equation}
 W_C[\mathscr{A}] = \int d\mu_{\Sigma}(g) \exp \left[  ig_{\rm YM} \int_{\Sigma: \partial \Sigma=C} F \right] ,
\end{equation}
where $d\mu_{\Sigma}(g):=\prod_{x \in \Sigma} d\mu(g_{x})$, with an invariant measure $d\mu$ on $G$  normalized as $\int d\mu(g_{x})=1$, $g_{x}$ is an element of a gauge group $G$ (more precisely, rep. $D_R(g_{x})$ of  $G$),
the two-form $F:=dA=\frac12 F_{\mu\nu}(x) dx^\mu \wedge dx^\nu$ is defined from the one-form $A := A_\mu(x) dx^\mu$, 
$
A_\mu(x) = {\rm tr}\{ \rho[ g_{x}^\dagger \mathscr{A}_\mu(x) g_{x} + ig_{\rm YM}^{-1} g_{x}^\dagger \partial_\mu g_{x} ] \} ,  
$ 
by
\begin{align}
  F_{\mu\nu}(x) &=  \sqrt{2(N-1)/N} [\mathscr{G}_{\mu\nu} (x)   
+ ig_{\rm YM}^{-1} {\rm tr} \{ \rho g_{x}^\dagger [\partial_\mu, \partial_\nu] g_{x} \} ],
\end{align}
with the field strength $\mathscr{G}_{\mu\nu}$ defined by
\begin{align}
 \mathscr{G}_{\mu\nu} (x) 
  &:=   \partial_\mu {\rm tr} \{ \mathbf{n}(x) \mathscr{A}_\nu(x) \} - \partial_\nu {\rm tr} \{ \mathbf{n}(x) \mathscr{A}_\mu(x) \} 
\nonumber\\& 
+ \frac{2(N-1)}{N} ig_{\rm YM}^{-1} {\rm tr} \{ \mathbf{n}(x) [\partial_\mu \mathbf{n}(x), \partial_\nu \mathbf{n}(x) ] \} 
 ,
\end{align}
and a normalized traceless field $\mathbf{n}(x)$ called the color field  
\begin{equation}
 \mathbf{n}(x) :=  \sqrt{N/[2(N-1)]} g_{x} \left[ \rho - \bm{1}/{\rm tr}(\bm{1}) \right] g_{x}^\dagger .
\end{equation}
Here $\rho$ is defined as $\rho :=  | \Lambda \rangle \langle \Lambda |$ using a reference state (highest or lowest  weight state of the rep.) $| \Lambda  \rangle$ making a rep. of the Wilson loop we consider. 
Note that ${\rm tr}(\rho) = \langle \Lambda | \Lambda \rangle = 1$ follows from the normalization of $| \Lambda \rangle$. 

Finally, the Wilson loop operator in the fundamental rep. of $SU(N)$ reads \cite{Kondo08}
\begin{align}
& W_C[\mathscr{A}] 
=  \int  d\mu_{\Sigma}(g)  \exp \left\{  ig_{\rm YM} (k, \Xi_{\Sigma}) + ig_{\rm YM} (j, N_{\Sigma}) \right\} ,
\label{NAST-SUN}
\nonumber\\
& k:=   \delta *f = *df, \quad j:=  \delta f , 
\quad
f:=  \sqrt{2(N-1)/N}  \mathscr{G}  ,
\nonumber\\
& \Xi_{\Sigma} :=  * d\Theta_{\Sigma} \Delta^{-1} = \delta *\Theta_{\Sigma} \Delta^{-1} , \
 N_{\Sigma} := \delta \Theta_{\Sigma} \Delta^{-1} ,
\end{align}
where two conserved currents,   ``magnetic-monopole current''   $k$ and  ``electric current'' $j$, are introduced, 
$\Delta:=d\delta+\delta d$ is the $D$-dimensional Laplacian, and $\Theta$ is an antisymmetric tensor of rank two  called the vorticity tensor:
$
 \Theta^{\mu\nu}_{\Sigma}(x) 
:=   \int_{\Sigma}  d^2S^{\mu\nu}(x(\sigma)) \delta^D(x-x(\sigma))  ,
$
which  has the support on the surface $\Sigma$ (with the surface element $dS^{\mu\nu}(x(\sigma))$) whose boundary is the loop $C$.
Incidentally, the last part $ig_{\rm YM}^{-1} {\rm tr} \{ \rho g_{x}^\dagger [\partial_\mu, \partial_\nu] g_{x} \}$ in $F$ corresponds to the Dirac string \cite{Kondo97,Kondo98a}, which is not gauge invariant and does not contribute to the Wilson loop in the end. 

For $SU(3)$ in the fundamental rep., the lowest-weight state $\langle \Lambda |=(0,0,1)$ leads to 
\begin{equation}
 \mathbf{n}(x) = g_{x} (\lambda_8/2) g_{x}^\dagger \in SU(3)/[SU(2) \times U(1)] \simeq CP^2 ,
\end{equation}
with the Gell-Mann matrix $\lambda_8:={\rm diag.}(1,1,-2)/\sqrt{3}$, 
while for $SU(2)$, $\langle \Lambda |=(0,1)$ yields
\begin{equation}
 \mathbf{n}(x) = g_{x} (\sigma_3/2) g_{x}^\dagger \in SU(2)/U(1) \simeq S^2 \simeq CP^1 ,
\end{equation}
with the Pauli matrix $\sigma_3:={\rm diag.}(1,-1)$.
The existence of magnetic monopole can be seen by a nontrivial Homotopy class of the map $\mathbf{n}$ from $S^2$ to the target space of the color field $\mathbf{n}$ \cite{KT99}:  
For $SU(3)$, 
\begin{align}
 & \pi_2(SU(3)/[SU(2) \times U(1)])=\pi_1(SU(2) \times U(1)) 
\nonumber\\&
=\pi_1(U(1))=\mathbb{Z} ,
\end{align}
while for $SU(2)$
\begin{equation}
 \pi_2(SU(2)/U(1))=\pi_1(U(1))=\mathbb{Z} .
\end{equation}
For $SU(3)$, the magnetic charge of the non-Abelian magnetic monopole obeys the quantization condition \cite{Kondo08}:
\begin{equation}
 Q_m := \int d^3x k^0 = 2\pi \sqrt{3} g_{\rm YM}^{-1} n , \ n \in \mathbb{Z} .
\end{equation}
The NAST shows that {\it the $SU(3)$ Wilson loop operator in the fundamental rep. detects the inherent $U(2)$ magnetic monopole which is  $SU(3)$ gauge invariant}. 
The rep. can be classified by its {\it stability group} $\tilde H$ of $G$ \cite{KT99,Kondo08}.  
For the fundamental rep. of $SU(3)$, the stability group  is $U(2)$.  
Therefore, the non-Abelian $U(2) \simeq SU(2)  \times U(1)$ magnetic monopole follows from  $\tilde H=SU(2)_{1,2,3} \times U(1)_{8}$, while the Abelian $U(1) \times U(1)$ magnetic monopole comes from $\tilde H=U(1)_{3} \times U(1)_{8}$. 
The adjoint rep. belongs to the latter case. 
The former case occurs only when the weight vector of the rep. is orthogonal to some of root vectors.  
The fundamental rep. is indeed  this case. 
For $SU(2)$, such a difference does not exist and $U(1)$ magnetic monopoles appear, since $\tilde H$ is always $U(1)$ for any rep.. 
For $SU(3)$, our result is different from Abelian projection: two independent $U(1)$ magnetic monopoles appear for any rep., since 
\begin{align}
 & \pi_2(SU(3)/U(1) \times U(1))=\pi_1(U(1) \times U(1)) =\mathbb{Z}^2 .
\end{align}

\section{Reformulating Yang-Mills theory using new variables}

For  $SU(3)$, two options  are possible, maximal for $\tilde H=U(1)^2$ \cite{Cho80c,FN99a} and minimal for $\tilde H=U(2)$ \cite{KSM08}. 
In the minimal one which gives the optimal description of quark in the fundamental rep., we consider the decomposition 
\begin{equation}
 \mathscr{A}_\mu(x) = \mathscr{V}_\mu(x) + \mathscr{X}_\mu(x) ,
 \label{decomp}
\end{equation}
such that (a)   $\mathscr{V}_\mu$ alone reproduces the Wilson loop operator: 
\begin{equation}
 W_C[\mathscr{A}] = W_C[\mathscr{V}] ,
 \label{W-dominant}
\end{equation}
and that (b) the field strength $\mathscr{F}_{\mu\nu}[\mathscr{V}] := \partial_\mu \mathscr{V}_\nu - \partial_\nu \mathscr{V}_\mu -ig_{\rm YM} [ \mathscr{V}_\mu,   \mathscr{V}_\nu ]$ in the color direction $\bm{n}$ agrees with $\mathscr{G}_{\mu\nu}$:
\begin{align}
 \mathscr{G}_{\mu\nu}(x) ={\rm tr} \{ \bm{n}(x) \mathscr{F}_{\mu\nu}[\mathscr{V}](x) \} .
 \label{G-NF}
\end{align}
The fields  $\mathscr V_\mu(x)$ and $\mathscr X_\mu(x)$ are  determined by solving defining equations, once the color field $\bm{n}(x)$ is given:
\\
\noindent
(I)  $\bm{n}(x)$ is a covariant constant in the background $\mathscr{V}_\mu(x)$:
\begin{align}
  0 = D_\mu[\mathscr{V}] \bm{n}(x) 
:=\partial_\mu \bm{n}(x) -  ig_{\rm YM} [\mathscr{V}_\mu(x), \bm{n}(x)]
 ,
\label{defVL2}
\end{align}
(II)  $\mathscr{X}^\mu(x)$  does not have the $\tilde{H}$-commutative part:
\begin{align}
  \mathscr{X}^\mu(x)_{\tilde{H}} := \left( {\bf 1} -   2\frac{N-1}{N}  [\bm{n} , [\bm{n} ,  \cdot ]]
\right) \mathscr{X}^\mu(x)   = 0  
\label{defXL2}
 . 
\end{align}
Indeed, (II) guarantees (a) and (I) guarantees (b). 
This is also checked by using the explicit form of decomposed fields which are uniquely fixed: 
\begin{align}
 \mathscr{X}_\mu  =& -ig_{\rm YM}^{-1}  \frac{2(N-1)}{N}  [\bm{n} , \mathscr{D}_\mu[\mathscr{A}]\bm{n}  ]
\in \mathscr{L}(G/\tilde{H}) ,
\nonumber\\
\mathscr V_\mu 
 =& \mathscr C_\mu 
  +\mathscr B_\mu 
 ,
\nonumber\\
  \mathscr{C}_\mu 
=& \mathscr{A}_\mu  - \frac{2(N-1)}{N}   [\bm{n} , [ \bm{n} , \mathscr{A}_\mu ] ]
\in  \mathscr{L}(\tilde{H}) 
,
\nonumber\\
 \mathscr{B}_\mu 
=& i g_{\rm YM}^{-1} \frac{2(N-1)}{N}[\bm{n} , \partial_\mu  \bm{n}  ] 
\in \mathscr{L}(G/\tilde{H}) 
 .
 \label{NLCV-minimal}
\end{align}

In our reformulation, $\mathscr{V}_\mu(x)$ and $\mathscr{X}_\mu(x)$ must be  expressed in terms of $\mathscr{A}_\mu(x)$.
Therefore, we must give a procedure of determining $\bm{n}$ from  $\mathscr A_\mu$, thereby, all the new variables  $\mathscr{C}_\mu$, $\mathscr{X}_\mu$ and $\bm{n}$  are obtained 
from   $\mathscr{A}_\mu$: 
\begin{equation}
 \mathscr{A}_\mu^A  \Longrightarrow (\bm{n}^\beta, \mathscr{C}_\nu^k,  \mathscr{X}_\nu^b)  
 .
\end{equation} 
We begin with counting degrees of freedom:
$\mathscr A_\mu \in \mathscr{L}(G)=su(N)$ 
 means  
$\#[\mathscr A_\mu^A]=D  \cdot  {\rm dim}G=D(N^2-1)$, 
$\mathscr C_\mu \in \mathscr{L}(\tilde{H})=u(N-1)$
means 
$
\#[\mathscr C_\mu^k]=D  \cdot  {\rm dim}\tilde{H}=D(N-1)^2 
$ 
and 
$\mathscr X_\mu \in \mathscr{L}(G/\tilde{H})$ 
means
$
\#[\mathscr X_\mu^b]
=D  \cdot  {\rm dim}(G/\tilde{H})=2D(N-1)   
$ 
and
$\bm{n}  \in \mathscr{L}(G/\tilde{H})$ 
means
$\#[\bm{n}^\beta]= {\rm dim}(G/\tilde{H})=2(N-1)
$. 
Thus, the new variables $(\bm{n}^\beta, \mathscr{C}_\nu^k,  \mathscr{X}_\nu^b)$ have the $2(N-1)$ extra degrees of freedom, to be eliminated   to obtain the new theory equipollent to the original one. For this purpose, we impose $2(N-1)$ constraints $\bm\chi=0$, which we call the reduction condition.
For example, minimize the functional
\begin{align}
R[\mathscr{A}, \bm{n} ]
 :=   \int d^Dx \frac12 (D_\mu[\mathscr{A}]\bm{n})^2,
\end{align}
with respect to the enlarged gauge transformation: 
$
\delta\mathscr{A}_\mu=D_\mu[\mathscr{A}]\bm\omega,
$
and
$
\delta \bm{n}
 =gi[\bm\theta , \bm{n} ]
 =gi[ \bm\theta_\perp , \bm{n} ]
$
where 
$\bm\omega \in \mathscr{L}(G)$ and $\bm\theta_\perp \in \mathscr{L}(G/\tilde{H})$.
Then, we find 
$
\delta R[\mathscr{A}, \bm{n}]
=g \int d^Dx
 (\bm\theta_\perp-\bm\omega_\perp)
 \cdot i[ \bm{n} , D^\mu[\mathscr{A}]D_\mu[\mathscr{A}]\bm{n} ]
 ,
$
where $\bm\omega_\perp$ denotes the component of $\bm\omega$ in the direction $\mathscr{L}(G/\tilde{H})$. 
The minimization  $\delta R[\mathscr{A}, \bm{n}]=0$ imposes no condition for $\bm\omega_\perp = \bm\theta_\perp$ (diagonal part of $G \times G/\tilde{H}$), while 
\begin{equation}
\bm\chi[\mathscr{A},\bm{n}]
 :=[ \bm{n} ,  D^\mu[\mathscr{A}]D_\mu[\mathscr{A}]\bm{n} ]
 = 0 
  ,
\label{eq:diff-red}
\end{equation}
is imposed for $\bm\omega_\perp   \not=  \bm\theta_\perp$ (off-diagonal part of $G \times G/\tilde{H}$).
The number of constraint is 
$
\#[\bm\chi]= {\rm dim}(G \times G/\tilde{H})- {\rm dim}(G)= {\rm dim}(G/\tilde{H})=2(N-1)=\#[\bm{h}^\beta]
$ as desired. 
As a bonus, the color field $\bm{n}(x)$ is determined by solving (\ref{eq:diff-red}) for  given   $\mathscr{A}_\mu(x)$.
This completes the procedure.

The Wilson loop average $W_{C}$ is defined by 
\begin{align}
 W_{C} =& Z_{{\rm YM}}^{-1} \int \mathcal{D}\mathscr{A}_\mu^A e^{-S_{{\rm YM}}[\mathscr{A}]} W_C[\mathscr{A}] ,
\end{align}
with  the partition function 
$
Z_{{\rm YM}} =  \int \mathcal{D}\mathscr{A}_\mu^A e^{-S_{{\rm YM}}[\mathscr{A}]} 
$
by omitting the gauge fixing to simplify the expression. 
The pre-NAST (\ref{pre-NAST}) tells us that
\begin{equation}
 W_{C} =  Z_{{\rm YM}}^{-1} \int  d\mu_{C}(g)   \mathcal{D}\mathscr{A}_\mu^A e^{-S_{{\rm YM}}[\mathscr{A}]} 
   e^{  ig_{\rm YM} \oint_{C} A } .
\end{equation}
Inserting  
$
1 = \int \mathcal{D}n^\alpha \prod_{x}  \delta(\bm{n}(x)-g_{x}(\lambda_8/2)g^\dagger_{x})
$ 
yields 
\begin{align}
 W_{C} =&  Z_{{\rm YM}}^{-1} \int  d\mu_{C}(g) \int \mathcal{D}\mathscr{A}_\mu^A  \mathcal{D}n^\alpha \delta(\bm{n}(x)-g_{x}(\lambda_8/2)g^\dagger_{x})
 \nonumber\\&
 \times   e^{-S_{{\rm YM}}[\mathscr{A}]}  e^{  ig_{\rm YM} \oint_{C} A  } .
\end{align}
Thus, in the reformulated theory in which $n^\beta(x)$, $\mathscr{C}_\nu^k(x)$, $\mathscr{X}_\nu^b(x)$ are {\it independent} field variables,  $W_{C}$ is written
\begin{align}
 W_{C} =&  \tilde{Z}_{{\rm YM}}^{-1} \int  d\mu_{\Sigma}(g) \int  \mathcal{D}\mathscr{C}_\nu^k \mathcal{D}\mathscr{X}_\nu^b  \mathcal{D}n^\beta
\delta(\tilde{\bm\chi}) \Delta_{\rm FP}^{\rm red}
\tilde{J}
 \nonumber\\&
 \times   e^{-\tilde S_{\rm YM}[\bm n, \mathscr{C},\mathscr{X}]}  e^{  ig_{\rm YM} (k, \Xi_{\Sigma}) + ig_{\rm YM} (j, N_{\Sigma}) } ,
\end{align}
where 
the Yang-Mills action is rewritten in terms of new variables using (\ref{decomp}) and (\ref{NLCV-minimal}), 
$
\tilde S_{\rm YM}[\bm n, \mathscr{C},\mathscr{X}]=S_{\rm YM}[\mathscr A]
$
and the new partition function is introduced:
$
 \tilde{Z}_{{\rm YM}}  
= \int \mathcal{D}\mathscr{C}_\nu^k \mathcal{D}\mathscr{X}_\nu^b  \mathcal{D}n^\beta
\delta(\tilde{\bm\chi}) \Delta_{\rm FP}^{\rm red} \tilde{J}
 e^{-\tilde S_{\rm YM}[\bm n, \mathscr{C},\mathscr{X}]} 
$.
It is shown \cite{KSM08} that the integration measure $\mathcal{D}\mathscr{A}_\mu^A$  is finally transformed to 
$
 \mathcal{D}\mathscr{C}_\nu^k \mathcal{D}\mathscr{X}_\nu^b \mathcal{D}n^\beta 
\delta(\tilde{\bm\chi}) \Delta_{\rm FP}^{\rm red} \tilde{J}
$, 
where 
(i) the Jacobian  $\tilde{J}$ 
is very simple, 
$
  \tilde{J} = 1 ,
$
\cite{KSM08} 
irrespective of the choice of reduction condition,  
(ii)  $\bm{\chi}[ \mathscr{A},\bm{n}]=0$ is rewritten in terms of new  variables:
$
\tilde{\bm\chi} 
 :=\tilde{\bm\chi} [\bm n, \mathscr{C},\mathscr{X}]
 :=D^\mu[\mathscr{V}]\mathscr{X}_\mu 
 , 
$
and 
(iii) the associated Faddeev-Popov determinant  $\Delta_{\rm FP}^{\rm red}$   
is calculable   using the BRST method, e.g.\cite{KMS05}.

\section{Numerical simulations}

The SU(3) Yang-Mills theory can be reformulated in the continuum and on a lattice using new variables. 
 For  $SU(3)$, two options  are possible, maximal for $\tilde H=U(1)^2$ \cite{Cho80c,FN99a} and minimal for $\tilde H=U(2)$ \cite{KSM08}. 
In our reformulation,  all the new variables  $\mathscr{C}_\mu$, $\mathscr{X}_\mu$ and $\mathbf{n}$  are obtained 
from   $\mathscr{A}_\mu$: 
\begin{equation}
 \mathscr{A}_\mu^A  \Longrightarrow (\mathbf{n}^\beta, \mathscr{C}_\nu^k,  \mathscr{X}_\nu^b)  
 ,
\end{equation} 
once the color field $\mathbf{n}$ is determined by solving the reduction condition:
\begin{equation}
\bm\chi[\mathscr{A},\mathbf{n}]
 :=[ \mathbf{n} ,  D^\mu[\mathscr{A}]D_\mu[\mathscr{A}]\mathbf{n} ]
 = 0 
  ,
\label{eq:diff-red}
\end{equation}

On a four-dimensional Euclidean lattice, gauge field configurations  $\{ U_{x,\mu} \}$ are generated by using the standard Wilson action and pseudo heat-bath method.  
For a given $\{ U_{x,\mu} \}$,   color field $\{ \bm{n}_{x} \}$ are determined by imposing a lattice version of  reduction condition.  Then new variables are introduced by using the lattice version of  change of variables \cite{lattice-f}. 

\begin{figure}[h]
\includegraphics[height=7.0cm,width=10.0cm]{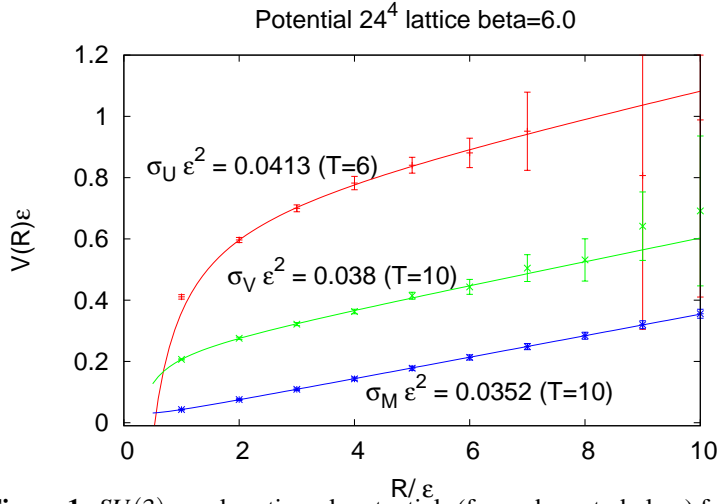}
\vspace{-0.6cm}
\caption{$SU(3)$ quark-antiquark potential: (from above to below) 
 full potential $V_f(r)$,  restricted part $V_a(r)$  and  magnetic--monopole part  $V_m(r)$  at $\beta=6.0$ on $24^4$ ($\epsilon$: lattice spacing).}
\label{fig:quark-potential}
\end{figure}

Fig. \ref{fig:quark-potential} shows the full $SU(3)$ quark-antiquark potential $V(r)$ obtained from the $SU(3)$ Wilson loop average $\langle W_C[\mathscr{A}] \rangle$, the restricted part $V_a(r)$ from the  $\mathscr{V}$ Wilson loop average $\langle W_C[\mathscr{V}] \rangle$, and  magnetic--monopole part  $V_m(r)$ from    $\langle e^{  ig_{\rm YM} (k, \Xi_{\Sigma})  }  \rangle$. They are gauge invariant quantities by construction.  These results exhibit infrared $\mathscr{V}$ dominance in the string tension (85--90\%) and  non-Abelian $U(2)$ magnetic monopole dominance in the string tension  (75\%) in the gauge independent way.

\begin{figure}[h]
\begin{center}
\includegraphics[height=5.0cm,width=7.0cm]{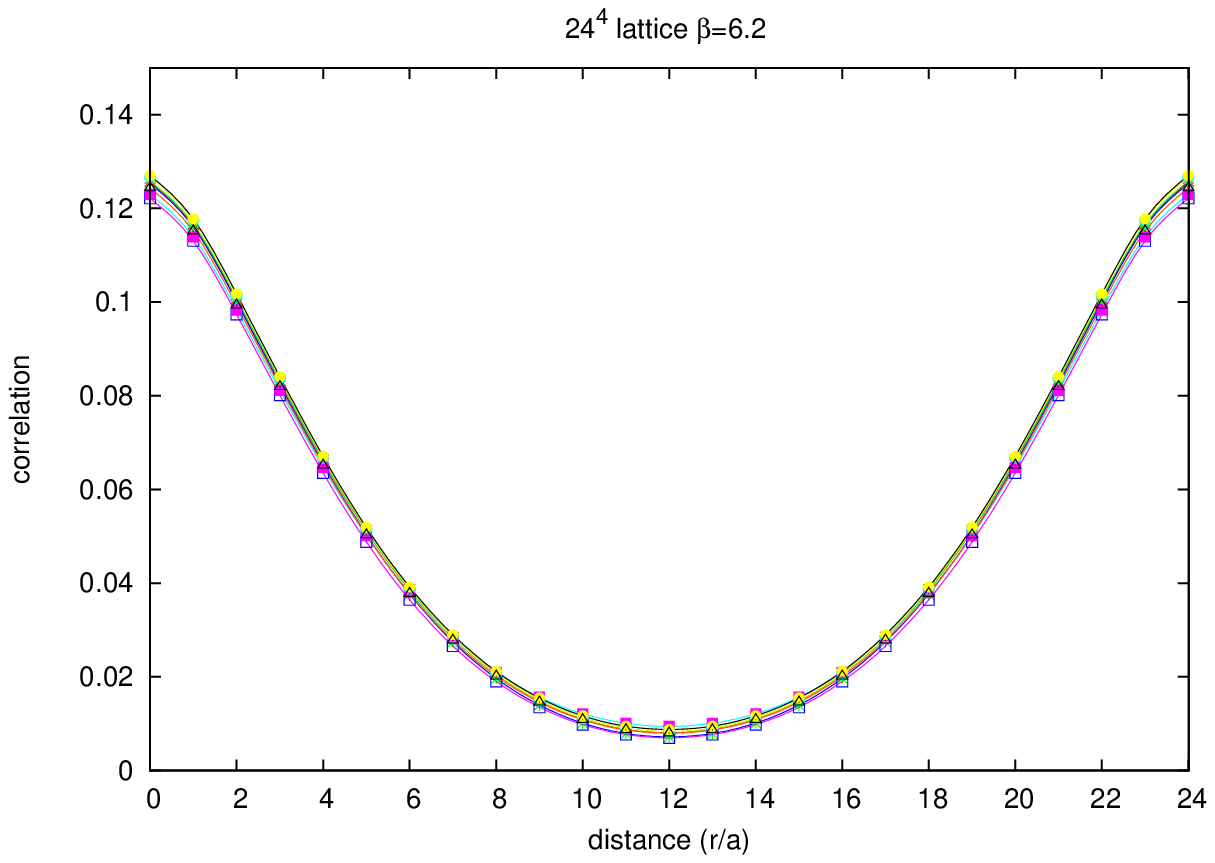}
\includegraphics[height=5.0cm,width=7.0cm]{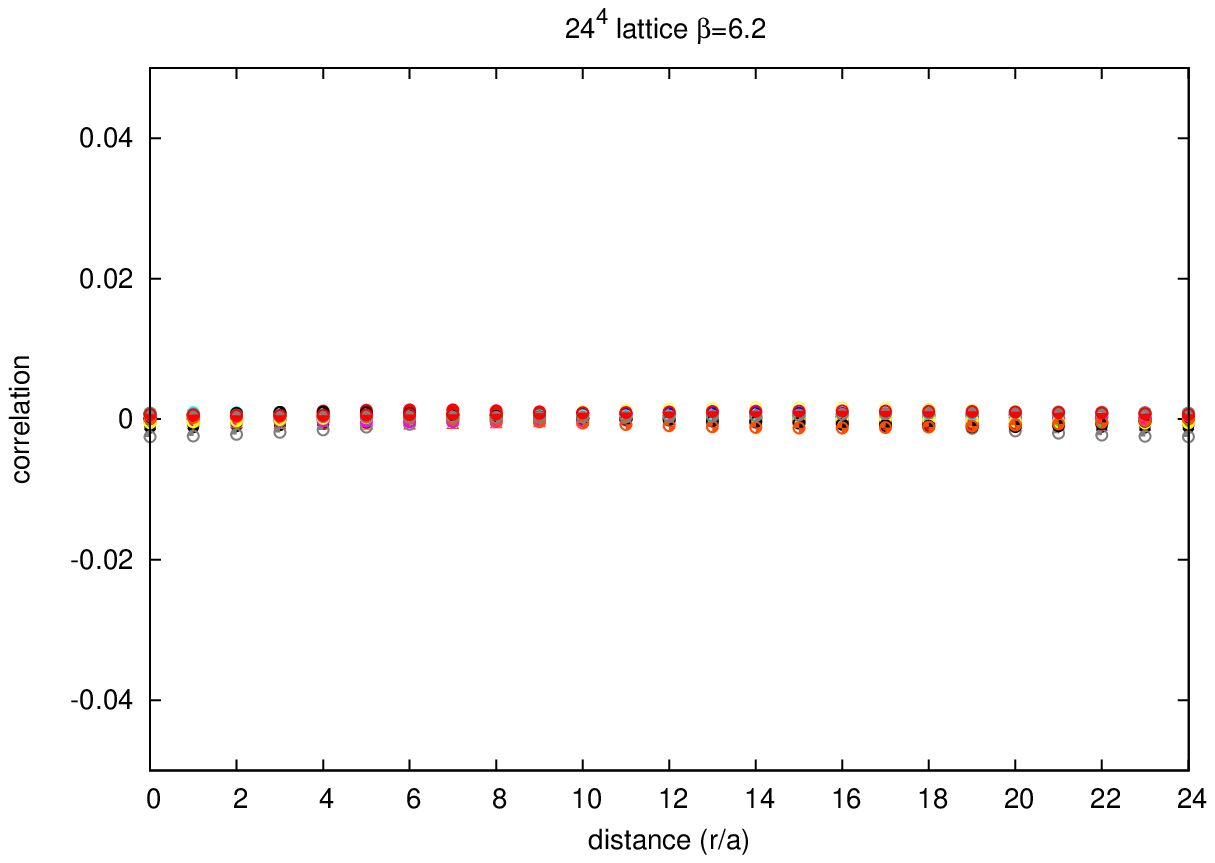}
\end{center}
\vspace{-0.6cm}
\caption{Color field correlators $\langle  n^A(0) n^B(r) \rangle$ ($A,B=1, \cdots, 8$) measured at  $\beta=6.2$ on   24$^4$ lattice,  using 500 configurations under the Landau gauge. 
(Left) $A=B$, 
(Right) $A \not= B$.
}
\label{fig:color-field-corr}
\end{figure}

To obtain correlation functions of field variables, we need to fix the gauge and we have adopted the Landau gauge. 
Fig.\ref{fig:color-field-corr} shows  two-point correlation functions of color field, indicating  the global $SU(3)$ {\it color symmetry preservation}, no specific direction in color space: $\langle  n^A(0) n^B(r) \rangle = \delta^{AB} D(r)$.
We have also checked that  one-point functions vanish, 
$\langle  n^A(x)  \rangle = \pm 0.002 \simeq 0$.

\begin{figure}[h]
\begin{center}
\includegraphics[height=6.0cm,width=10.0cm]{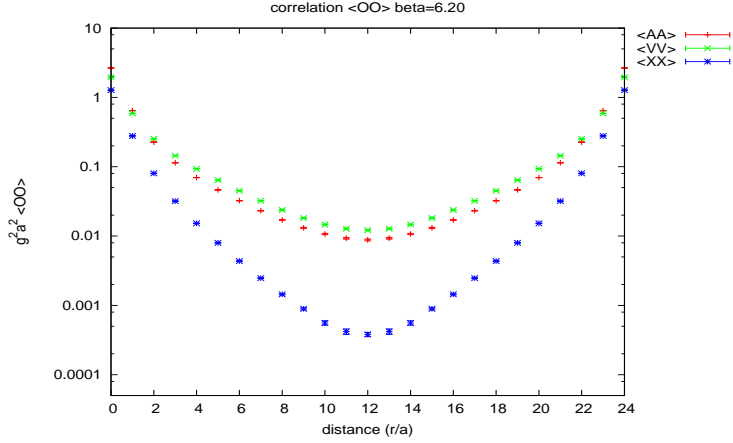}
\end{center}
\vspace{-0.6cm}
\caption{Field correlators as functions of $r$ (from above to below)
$\langle  \mathscr{V}_\mu^A(0) \mathscr{V}_\mu^A(r) \rangle$, 
$\langle  \mathscr{A}_\mu^A(0) \mathscr{A}_\mu^A(r) \rangle$, 
and
$\langle  \mathscr{X}_\mu^A(0) \mathscr{X}_\mu^A(r) \rangle$.
}
\label{fig:decomp-field-corr}
\end{figure}

Fig. \ref{fig:decomp-field-corr} shows correlators of new fields $\mathscr{V}$, $\mathscr{X}$, and original fields $\mathscr{A}$, indicating the {\it infrared  dominance} of restricted correlation functions in the sense that the  variable $\mathscr{V}$ is dominant in the long distance, while the correlator of $SU(3)/U(2)$ variable $\mathscr{X}$  decreases quickly.  
For  $\mathscr{X}$,  at least, we can introduce a gauge-invariant mass term $\frac12 M_X^2 \mathscr{X}_\mu^A \mathscr{X}_\mu^A $, since $\mathscr{X}$ transforms like an adjoint matter field under the gauge transformation. 
The naively estimated ``mass" of $\mathscr{X}$ is 
$M_X = 2.409 \sqrt{\sigma_{\rm phys}} = 1.1$ GeV.  
 This value should be compared with the result in MA gauge. 
The details of numerical results will be given in a subsequent paper. 
These results give  numerical evidences for non-Abelian dual superconductivity as a mechanism for quark confinement in SU(3) Yang-Mills theory.


\section{Conclusion}

We have shown: 
(i) The $SU(N)$ Wilson loop operator can be rewritten in terms of a pair of gauge-invariant magnetic-monopole current  $k$ ($(D-3)$-form) and the associated geometric object defined from the Wilson surface $\Sigma$ bounding the Wilson loop $C$, and another pair of an electric current $j$ (one-form  independently of $D$) and the associated topological object, which follows from a non-Abelian Stokes theorem for the Wilson loop operator \cite{Kondo08}.   
(ii) The $SU(N)$ Yang-Mills theory can be reformulated in terms of new field variables obtained by change of variables from the original Yang-Mills gauge field $\mathscr{A}_\mu^A(x)$ \cite{KSM08}, so that it gives an optimal description for the non-Abelian magnetic monopole defined from the $SU(N)$ Wilson loop operator in the fundamental rep. of quarks.
(iii) A lattice version of the reformulated  Yang-Mills theory can be constructed \cite{lattice-f}.
Numerical simulations of the  lattice $SU(3)$ Yang-Mills theory give numerical evidences that the restricted field variables become dominant in the infrared for correlation functions and the string tension ({\it infrared restricted non-Abelian dominance}) and that the $U(2)$ magnetic monopole gives a most dominant contribution to the string tension obtained from  $SU(3)$ Wilson loop average ({\it non-Abelian magnetic monopole dominance}). 
See \cite{KSSK10} for more informations.

\section*{Acknowledgements} 
The author wishes to thank the organisers for the invitation to the very pleasant meeting in such a wonderful city.
He would also like to thank all the participants
of the workshop gThe many faces of QCDh for interesting discussions.
This work is  supported by Grant-in-Aid for Scientific Research (C) 21540256 from Japan Society for the Promotion of Science
(JSPS).

\end{document}